\documentclass[12pt]{article}
\usepackage[a4paper, total={6in, 10in}]{geometry}
\usepackage[
natbib,
backend=biber,
style=apa,
sorting=ynt
]{biblatex}
\usepackage{graphicx} 
\usepackage[acronym,nohypertypes={acronym,notation}]{glossaries}
\usepackage{xcolor}
\usepackage{booktabs}
\usepackage{amssymb}
\usepackage{pifont}
\usepackage{multirow}
\usepackage{enumitem}
\newcommand{\cmark}{\ding{51}}%
\newcommand{\xmark}{\ding{55}}%
\definecolor{pers_red}{HTML}{B85450}
\definecolor{pers_green}{HTML}{82B366}

\glsdisablehyper

\title{Estimation of Head Motion in Structural MRI and its Impact on Cortical Thickness Measurements in Retrospective Data} 

\author{Charles Bricout$^{1}$,
Samira Ebrahimi Kahou$^{2,3}$,
Sylvain Bouix$^{1\ast}$,
\\
{\small $^{1}$École de technologie supérieure, Montréal, Canada,}\\
{\small $^{2}$University of Calgary, Calgary, Canada}\\
{\small $^{3}$Canada CIFAR AI Chair/Mila}\\
{\small $^\ast$Correspondence:  sylvain.bouix@etsmtl.ca}
}

\begin{document} 
\newacronym{mri}{MRI}{Magnetic Resonance Imaging}
\newacronym{t1w}{T1w}{T1-weighted}
\newacronym{qc}{QC}{Quality Control}
\newacronym{vnav}{vNavs}{Volumetric Navigators}

\newacronym{hbn}{HBN}{Healthy Brain Network}
\newacronym{hcp}{HCP}{Human Connectome Project}
\newacronym{hcpep}{HCP-EP}{Human Connectome Project Early Psychosis}
\newacronym{hcpya}{HCP-YA}{Human Connectome Project Yound Adult}
\newacronym{abcd}{ABCD}{Adolescent Brain Cognitive Development}
\newacronym{mrart}{MR-ART}{Movement-Related Artefacts}

\newacronym{asd}{ASD}{Autism Spectrum Disorder}
\newacronym{adhd}{ADHD}{Attention Deficit Hyperactivity Disorder}

\newacronym{rubic}{RUBIC}{Rutgers University Brain Imaging Center}
\newacronym{cbic}{CBIC}{CitiGroup Cornell Brain Imaging Center}
\newacronym{cuny}{CUNY}{City College of New York}

\newacronym{rms}{RMS}{Root Mean Square}
\newacronym{rmse}{RMSE}{Root Mean Square Error}
\newacronym{ssim}{SSIM}{Structural Similarity Index Measure }

\newacronym{sfcn}{SFCN}{Simple Fully Convolutional Network}
\newacronym{cnn}{CNN}{Convolutional Neural Network}
\newacronym{kl}{KL}{Kullback-Leibler}
\newacronym{glm}{GLM}{Generalized Linear Model}

\newacronym{fdr}{FDR}{False Discovery Rate}
\newacronym{icc}{ICC}{Intraclass Correlation Coefficient}
\maketitle 


\begin{abstract}
   Motion-related artifacts are inevitable in Magnetic Resonance Imaging (MRI) and can bias automated neuroanatomical metrics such as cortical thickness. These biases can interfere with statistical analysis which is a major concern as motion has been shown to be more prominent in certain populations such as children or individuals with ADHD. Manual review cannot objectively quantify motion in anatomical scans, and existing quantitative automated approaches often require specialized hardware or custom acquisition protocols. Here, we train a 3D convolutional neural network to estimate a summary motion metric in retrospective routine research scans by leveraging a large training dataset of synthetically motion-corrupted volumes. We validate our method with one held-out site from our training cohort and with 14 fully independent datasets, including one with manual ratings, achieving a representative $R^2 = 0.65$ versus manual labels and significant thickness-motion correlations in 12/15 datasets. Furthermore, our predicted motion correlates with subject age in line with prior studies. Our approach generalizes across scanner brands and protocols, enabling objective, scalable motion assessment in structural MRI studies without prospective motion correction. By providing reliable motion estimates, our method offers researchers a tool to assess and account for potential biases in cortical thickness analyses.
\end{abstract}

\smallskip \noindent \textbf{Funding :} This research is supported by the National Institute of Mental Health (U24MH124629), Canada Research Chairs Program,  Canada Foundation for Innovation (SB), CIFAR (SEK) and the Digital Research Alliance of Canada.

\smallskip \noindent \textbf{Data Availability :} This research used retrospective human subject data made available by the Human Connectome Project, the Child Mind Institute and the OpenNeuro platform. Approval was granted by the Research Ethics Committee of École de technologie supérieure. 

\section{Introduction}

Motion artifacts are an inherent challenge in \gls{mri}. \cite{zaitsev_motion_2015} explain that, as a single structural MRI acquisition lasts approximately five minutes, involuntary motion is inevitable. Even more problematic, this motion can vary widely in intensity, from undetectable to full corruption, and can manifest as blurring, ghosting, and fine concentric arcs. In fact, \cite{andre_toward_2015} estimated that motion artifacts cost on average \$115,000 per scanner per year. Multiple methods exist to reduce motion artifacts prospectively; \cite{tisdall_volumetric_2011} proposed a technique called \gls{vnav} relying on the acquisition of a low-resolution 3D volume throughout the acquisition sequence to compute and correct patient motion dynamically. Such methods, however, rely on a specific sequence and have a low temporal rate, rendering them unable to detect motion that occurs between navigator acquisitions. \cite{zaitsev_motion_2015} estimates that this issue  cannot be overcome solely through hardware improvements or a single methodological solution.

One of the multiple problematic aspects caused by motion is its impact on automatic measurement tools.
\cite{blumenthal_motion_2002} show that increasing levels of motion artifact, graded by manual labeling on a four-grade scale, are linked to a decrease in total gray matter volume estimation.  \cite{reuter_head_2015} demonstrate this effect further by using several anatomical analysis packages, including FreeSurfer, a standard tool for automatic neuroanatomical computation \citep{fischl_freesurfer_2012}. They show that increasing motion, estimated with \gls{vnav}, leads to decreased gray matter volume and cortical thickness estimations. Furthermore, \cite{alexander-bloch_subtle_2016} find that there may be systematic effects of subject motion even on good quality volumes. They estimate the tendency of a subject to move by computing the average motion during an fMRI sequence and comparing it to estimates of cortical thickness, cortical gray matter volume, and mean curvature. They find that, as motion increases, thickness and volume decrease, whereas mean curvature increases. They also demonstrate that the effect of motion is not uniform throughout the brain by studying the thickness of the four separate lobes.

There is significant evidence that motion artifacts impact automatic anatomical measurement. This issue becomes more problematic as we explore differences in the tendency for motion between populations. \cite{pardoe_motion_2016} investigate if motion is related to diagnosis, age, and gender in a cohort of 2141 subjects. They also estimate motion through fMRI and find a significantly increased tendency to move in subjects with \gls{asd}, \gls{adhd}, and schizophrenia; they also find similar tendencies in younger patients. Finally, they reinforce findings from \cite{alexander-bloch_subtle_2016} by finding a significant relationship between fMRI-estimated motion and multiple anatomical metrics, such as gray matter volume and frontal, temporal, and parietal mean curvature, even when considering volumes that are manually assessed to be free of artifacts. These findings are very important as they may question results on those populations that do not take motion into account in statistical analyses.

Hence, there is a strong need for motion estimation techniques. First, \cite{rosen_quantitative_2018} show a strong agreement between the Euler number, a cortical surface-based quality metric, and manual scoring of motion. They also show a significant relationship with the thickness of different regions of the brain. Although interesting, the Euler number is hardly interpretable, and manual scoring of motion is known to be noisy. Furthermore, their method is tested for generalizability on just one dataset, which might not be enough to guarantee robustness. On the other hand, \cite{pollak_estimating_2023} use a depth camera mounted on an \gls{mri} to accurately quantify the motion of 500 patients. A software approximates transformation matrices between successive head positions and computes the \gls{rms} deviation, a metric proposed by \cite{jenkinson_measuring_nodate}. They then train a SFCN-derived model to quantify motion, reaching an R² of 0.433 on the test set, and also found a significant correlation with subjects' age. Unfortunately, the test set contains only 75 subjects and they did not use an external dataset to assess generalizability. This data labeling method, while accurate, demands specific materials and knowledge and can hardly be scaled for large data acquisition projects.

To address the manual labeling problem, several methods have turned to synthetic data generation. \cite{mohebbian_classifying_2021} use synthetic motion to corrupt images and label the created motion by computing the Frobenius norm of displacement and rotation applied. They then determine five bins of motion severity based on this norm and train an ensemble of \glspl{cnn}, one for T1 and one for T2, to classify volumes. This ensemble model reaches an accuracy of 90.3\% on the original dataset and 78.2\% on an independent dataset. While the model is very accurate on synthetic data, it is not tested on real motion, and there is no proof of generalization to real data. It is also focused on 2D images, which might miss some artifacts. Finally, \cite{sciarra_reference-less_2022} use a very similar approach, applying a ResNet on the \gls{ssim} between original data and corrupted data, grouped into 10 severity classes. They train on four datasets from different sites, reaching 89\% accuracy on the test set. We identify similar limitations: they do not leave one dataset out for generalizability testing, models are only tested on synthetic artifacts, and the method uses 2D slices.

To address these limitations, we expand upon previous work \citep{bricout_improving_2025}. In this paper, we will (1) train a 3D \gls{sfcn} on simulated motion artifacts quantified with \gls{rms} deviation, (2) extensively test our motion regressor on 15 real datasets, searching for known correlations with cortical thickness estimation, and (3) compare our predicted score with manual labeling on real data.

\section{Method}

\subsection{Training Datasets}

We train our model using \gls{mri} from the \gls{hbn} dataset \citep{alexander_open_2017}. \gls{hbn} is an initiative to acquire and share a biobank of data on 10,000 young New York area subjects, with ages ranging from 5 to 21. We use data acquired at three different sites. Data acquired at \gls{rubic} use a 3.0 T Siemens Tim Trio scanner, whereas data from \gls{cuny} and \gls{cbic} use a 3.0 T Siemens Prisma scanner; all use a 32-channel coil. The acquisition protocol for \gls{t1w} volumes at all sites is derived from the \gls{hcp} project \citep{marcus_human_2013}. Additionally, some of the volumes acquired at \gls{cbic} and \gls{cuny} also include \gls{t1w} data with a protocol using \gls{vnav}, derived from the \gls{abcd} study \citep{tisdall_volumetric_2011}. This \gls{vnav} acquisition is designed to reduce the effects of motion; for this reason, we decide to use volumes from \gls{cbic} and \gls{cuny} to generate synthetic motion data. We use \gls{rubic} to assess the relationship between motion and cortical thickness.

As our training relies on simulated artifacts, we need to ensure that the volumes used for the synthetic generation process are as motion-free as possible. This helps reduce noise in our training data by making our synthetic labels as accurate as possible. We created a simple web-based tool to review the volumes from each site with a pass or fail rating. A volume would fail if any sign of motion could be seen on one of the three slices sampled from the volume.

We later improved this rating system with six possible labels: "Clean", "Barely Noticeable", "Noticeable", "Strong", "Unusable", and "Corrupted", to allow filtering on a finer level if the quantity of "Clean" and "Barely Noticeable" data were insufficient. We rate only the \gls{cuny} dataset using the six-point \gls{qc} scale and retain only volumes rated as "Clean" or "Barely Noticeable" for the next step. In total, we retain only 449 volumes out of the 4,079 available.

All volumes were reviewed by Rater One (CB). To assess reliability, we asked two additional independent raters to grade the same 50 randomly selected volumes. Below, we provide information about motion artifact effects and our grading scale:
\begin{enumerate}[itemsep=0mm]
    \item "Clean": no doubt about data quality  
    \item "Barely Noticeable": unsure but no clear effects  
    \item "Noticeable": clear lines or blurring, noisy white matter  
    \item "Strong": strong lines and blurring, unclear delimitation between gray and white matter  
    \item "Unusable": hard to distinguish any information  
    \item "Corrupted": corruption unrelated to motion (truncation, metal artifacts, etc.)  
\end{enumerate}

Representative examples for each label are also provided (Appendix Figure \ref{fig:appendix:motsample}). Inter-rater reliability is assessed using the \gls{icc} \citep{shrout_intraclass_1979}. We employ a two-way random-effects, average-measure, absolute-agreement model (ICC(2,k)). We use the same metric for intra-rater reliability by asking Rater One to grade the 50 selected samples six months after the initial grading. Overall, we observe significant inter- and intra-rater reliability, confirming the reproducibility of our volume selection method (Table \ref{tab:method:interintra}).

\begin{table}[]
    \centering
    \begin{tabular}{c c c c}
    \toprule
        & ICC(2,k) & P-Value & 95$\%$ CI\\
        \midrule
        Intra-Rater & 0.903 & 3.46e-28 & [0.8, 0.95] \\
        Inter-Rater & 0.883 & 1.44e-13 & [0.78, 0.94] \\
        \bottomrule
    \end{tabular}
    \caption{Inter-/Intra-rater reliability with \gls{icc}}
    \label{tab:method:interintra}
\end{table}

\subsection{Evaluation Datasets}

As our model is trained on synthetic artifacts only and on one dataset, we select a large number of datasets with varying protocols, hardware and populations intended for evaluating  the model generalisabiliy.

\textbf{\gls{mrart}} is a dataset developed specifically to study the impact of motion on brain \gls{mri} \citep{adam_narai_movement-related_2022}. It contains paired motion-free and motion-corrupted T1-weighted \gls{mri} scans from 148 healthy adults (ages 18–75). Structural volumes are acquired on a Siemens MAGNETOM Prisma 3T scanner with a 20-channel coil using a classic MPRAGE sequence with GRAPPA acceleration. For each participant, three scans were acquired with varying levels of motion:
\begin{itemize}
    \item STAND: subjects have to stay still.
    \item HM1: subjects are instructed to nod their head five times when signaled.
    \item HM2: subjects are instructed to nod their head 10 times when signaled.
\end{itemize}

Then, two neuroradiologists with over ten years of experience independently rated the clinical usability of all scans on a three-point scale:
\begin{itemize}
    \item Good (1): Diagnostically usable.
    \item Medium (2): Partial artifacts, limited clinical utility.
    \item Bad (3): Severe artifacts, unusable for diagnostics.
\end{itemize}
The raters were blind to the acquisition condition (STAND/HM1/HM2). To ensure consistency, they first harmonized their criteria on 100 independent scans and resolved ambiguities through consensus during the labeling process.

\textbf{\gls{hcpya}} provides high-quality T1w brain \gls{mri} for 1,200 healthy adults (ages 22–35) \citep{van_essen_human_2012}. This study uses a modified Siemens Skyra (3T) scanner with a stronger gradient achieved by using an enhanced gradient coil designed for 7T scanners and a 64-channel coil. For the acquisition sequence, a custom version of MPRAGE with GRAPPA acceleration is used. This, combined with a thorough quality control process, ensures high-quality volumes.

\textbf{\gls{hcpep}} focused on studying the early stages of schizophrenia  \citep{jacobs_introduction_2024}. The project recruited 303 participants aged 16–35, of which we use 299, we get a total of 383 high-quality T1-weighted brain \gls{mri}, as some participants were scanned multiple times. Participants were scanned across three different sites, each using a Siemens MAGNETOM Prisma 3T scanner with either a 32- or 64-channel coil and an MPRAGE sequence similar to \gls{hcp}. It also provides a general quality score between one (worst) and four (best) for most volumes.

\textbf{Auxiliary test datasets from OpenNeuro:} We use OpenNeuro to find other suitable datasets to test our models. Our criteria are:
\begin{itemize}
    \item Includes at least one subject between 5 and 21.
    \item Field strength should be three Tesla.
    \item Acquisition protocol should be similar to MPRAGE.
    \item No prospective motion correction techniques (e.g., \gls{vnav}).
\end{itemize}
The final dataset consists of 847 volumes gathered from 11 studies (Table \ref{tab:dataset:openneuro}). We have access to a variety of scanners; we are especially interested in ds000144 and ds003568 as they do not use Siemens hardware. ds005234 is also of interest as it uses a different Siemens scanner. While all studies use a sequence similar to MPRAGE, none employ identical parameters, which should also provide important variability for evaluating our method . We do not report these parameters as only partial information is available for each dataset, thus preventing direct comparison.

\begin{table}[h]
    \fontsize{9pt}{9pt}\selectfont
    \centering
    \begin{tabular}{lcccc}
    \toprule
      Dataset & \# Volumes & Age Range & Population & Scanner \\
    \midrule
    ds000115 & 99 & 11-30 & Healthy / Schizophrenic & Siemens Magnetom TrioTim\\
    ds000144 & 45 & 6-10 & Anxious Children & GE Discovery MR750 / Signa Excite\\
    ds000256 & 24 & 5-15 & Healthy & Siemens Magnetom TrioTim\\
    ds001486 & 195 & 8-15 & Healthy & Siemens Magnetom TrioTim\\
    ds001748 & 62 & 10-35 & Healthy & Siemens Magnetom TrioTim / Prisma\\
    ds002424 & 79 & 8-12 & Healthy / ADHD & Siemens Magnetom TrioTim\\
    ds002862 & 71 & 8-10 & Dyslexia & Siemens Magnetom Skyra\\
    ds002886 & 56 & 8-15 & Healthy & Siemens Magnetom TrioTim\\
    ds003499 & 93 & 8-25 & Healthy & Siemens Magnetom Prisma\\
    ds003568 & 49 & 12-19 & Healthy / Depression & GE Discovery MR750\\
    ds005234 & 74 & 6-13 & Healthy / Autism & Siemens Magnetom Verio\\
    \bottomrule
    \end{tabular}
    \caption{Datasets retrieved from OpenNeuro}
    \label{tab:dataset:openneuro}
    
\end{table}

\subsection{Data Processing}

\subsubsection{Pre-Processing}
For pre-processing, we use Clinica's t1-linear pipeline \citep{routier_clinica_2021}. First, bias field correction is applied using the N4ITK method \citep{tustison_n4itk_2010}. Next, an affine registration is performed using ANTs  \citep{avants_insight_2014} to align each image to the MNI space with the ICBM 2009c nonlinear symmetric template \citep{fonov_unbiased_2011}. This ensures data quality and that each brain is centered and approximately the same size.

Finally, we use Freesurfer v 7.4.1 with the default configuration on all the preprocessed volumes to compute the thickness of 34 different structures and an aggregated mean cortical thickness for each subject \citep{fischl_freesurfer_2012}. For this study, we only use the measurements from the left hemisphere. 

\subsubsection{Synthetic Motion Generation}
To create our synthetic data, we apply random synthetic motion using TorchIO \citep{perez-garcia_torchio_2021}. This transformation samples $N$ transformation matrices representing subject motion, constrained by maximum translation and rotation parameters, and concatenates their k-space into a final, corrupted k-space \citep{shaw_mri_2019}. Using the successive transformation matrices, we can quantify the motion using the \gls{rms} deviation \citep{jenkinson_measuring_nodate}. 
It is calculated as follows: let \( T_1 \) and \( T_2 \) be two transformation matrices, and let \( x_c \in \mathbb{R}^3 \) be the center of the \gls{mri} volume. We define \( A \) and \( t \) as:

\begin{align*}
M &= T_2 T_1^{-1} - I \\
  &= \begin{bmatrix}
      A & t \\
      0\,0\,0 & 0
     \end{bmatrix}.
\end{align*}

Given an estimation of the brain radius $R_c$, the \gls{rms} deviation is then given by:

\begin{equation}
E_{RMS} = \sqrt{ \frac{1}{5} R_c^2 \, \mathrm{Trace}(A^\top A) + (t + A x_c)^\top (t + A x_c) }.
\end{equation}

To ensure a uniform distribution of motion data, we sample an expected motion magnitude from a uniform distribution, $e_{\text{motion}}\sim\mathcal{U}(0.01, 4.0)$. Using the expected motion, we sample reasonable restrictions for maximum rotation and translation from the following distributions: $c_{\text{translation}}\sim\mathcal{U}(0, \max(e_{\text{motion}}, 1))$ and $c_{\text{rotation}}\sim\mathcal{U}(0, \max(2\times e_{\text{motion}}, 1))$. These values are found empirically. With our constraints, we can finally sample a full transformation and compute the \gls{rms} deviation. As each transform is randomly generated to roughly match a predetermined motion score, we defined a tolerance parameter of 0.02 to control whether we need to sample a new transformation; we repeat this process until we obtain a transformation that satisfies the tolerance.

To augment the anatomical variety of our dataset, we apply random sagittal plane flipping and elastic deformation before generating the synthetic motion.
We generate 300 synthetic volumes for each original clean volume, resulting in 137,700 synthetic volumes that we split for train, validation, and test in a 60-20-20 split. To study the effect of synthetic motion on cortical thickness, we also generate 50 samples for 20 random subjects of the selected volumes with elastic transformation and random flip disabled, resulting in 1020 volumes when accounting for the original volumes. We process all volumes with Freesurfer to compute the mean cortical thickness.

\subsubsection{Training Time Augmentation}

As different datasets can have widely different contrasts and artifact levels, we use TorchIO to simulate Gaussian noise, Gaussian blur, and Gamma corrections \citep{perez-garcia_torchio_2021}. As we aim to make our model as robust as possible to different datasets, we strive for broad contrast simulation and also use MONAI's random histogram shift \citep{cardoso_monai_2022}. We also apply a random flip along the sagittal plane.

For a given volume, we separate our augmentation pipeline into two stages:
\begin{itemize}
    \item Base augmentation: we apply a flip with a probability of 0.5 and either add random Gaussian noise or apply smoothing with a probability of 0.8.
    \item Contrast augmentation: we apply either random gamma changes or a histogram shift on every volume.
\end{itemize}
For each stage, we randomly apply one of the proposed transformations to avoid augmenting a volume to the point of corruption.

Finally, to reduce the computational load inherent to 3D \glspl{cnn}, we crop our volume to a ROI of (160, 192, 160) and, to ensure consistency between volumes, we scale voxel intensities to the (0,1) range using min-max scaling.

\begin{figure}
    \centering
    \includegraphics[width=\linewidth]{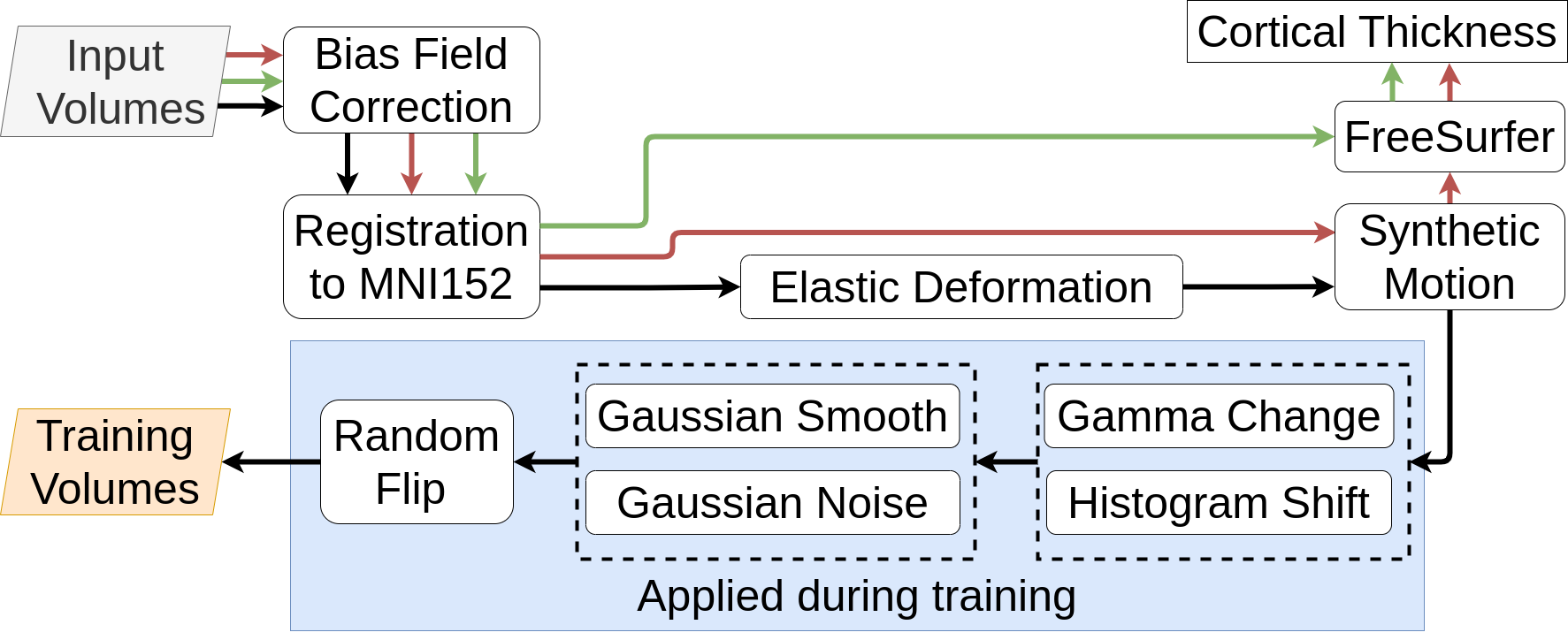}
    \caption{Data processing flow. \textbf{Black arrows} represents training data, \textcolor{pers_red}{\textbf{red arrows}} represents data used to assess synthetic motion effect, and \textcolor{pers_green}{\textbf{green arrows}} represents test data from real datasets.}
    \label{fig:enter-label}
\end{figure}

\subsection{Neural Networks Architectures}

\begin{figure}
\includegraphics[width=\linewidth]{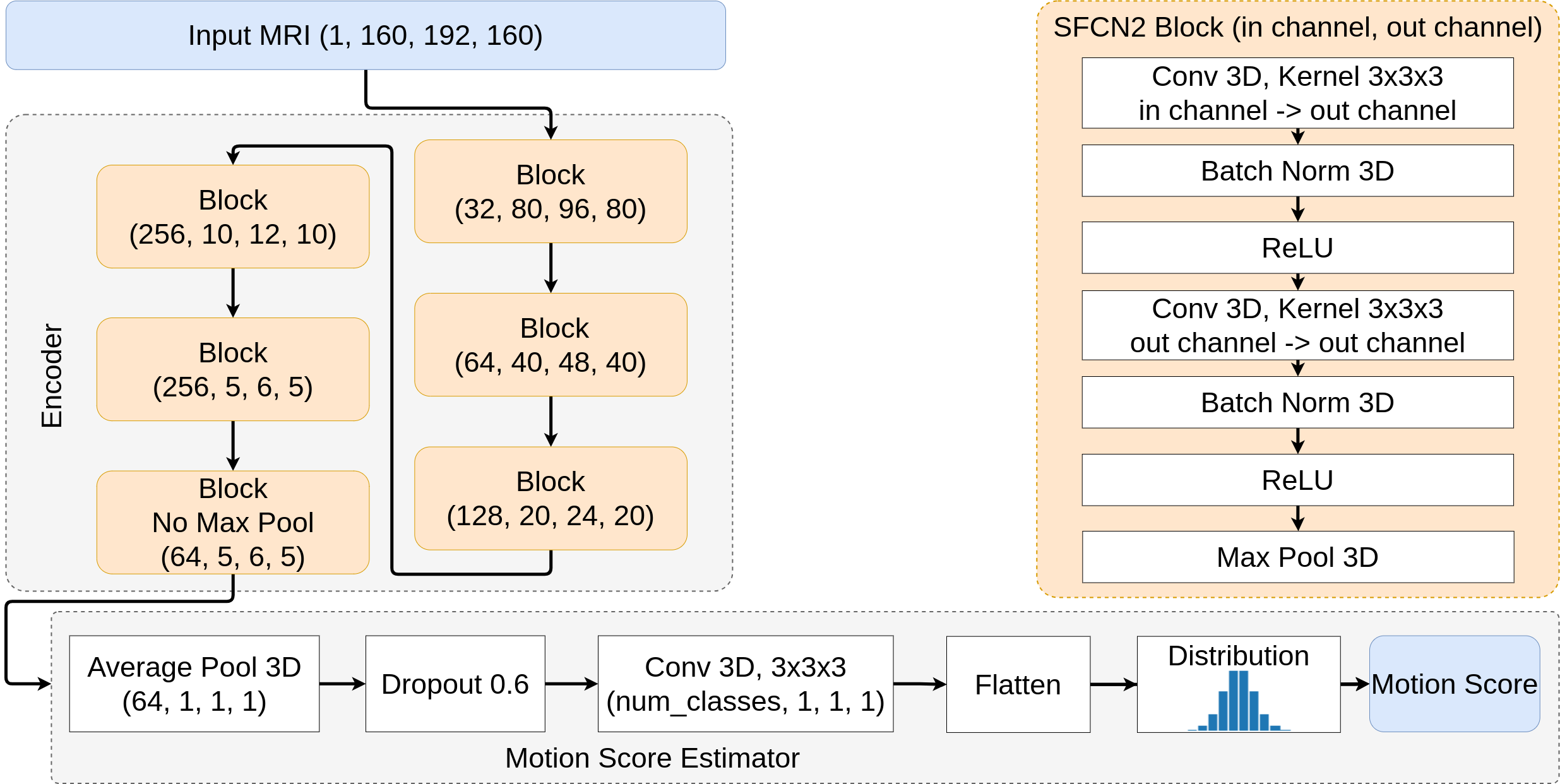}
    \caption{Description of the SFCN2 architecture}
    \label{fig:networks:sfcn}
\end{figure}

Expanding on our previous study on synthetic motion prediction, we use the \gls{sfcn} (Figure \ref{fig:networks:sfcn}) as our baseline architecture \citep{bricout_improving_2025}. \gls{sfcn} is a lightweight 3D \gls{cnn} architecture first proposed for brain age prediction \citep{peng_accurate_2021} and then used for head motion estimation \citep{pollak_estimating_2023}. 

Instead of directly regressing the motion score, we define a range of possible motion [-0.5,4.5] that we discretize in $N_{\text{bins}}=50$. Then, the network learns to predict a distribution over these bins, representing the probability of the noise magnitude. Finally, we train the \gls{cnn} by minimizing the \gls{kl} divergence between the predicted distribution and a normal distribution centered on the true motion score. 

Our model's encoder consists of six blocks, with five blocks performing downsampling and a sixth block adding more non-linearity and parameters to our network without changing the shape of our data. We also use the proposed channel configuration for each layer: [32, 64, 128, 256, 256, 64, 40]. For  motion inference, we compute a weighted average of each motion bin: 

$$\text{motion score}=\sum_{i}^{N_{\text{bins}}}x_i\cdot\text{motion}_i$$
where $x_i$ is the predicted probability for the $i$-th bin, and $\text{motion}_i$ is the center value of the $i$-th bin. 

Our model is a slightly heavier variation of \gls{sfcn}, using two convolutions per block instead of one. This architecture is more expressive, which might help capture subtlety while still being light enough for efficient training. We refer to this network as SFCN2.

\subsection{Training}

We train our models using the AdamW optimizer with a learning rate of 0.001 and a weight decay of 0.1, both selected empirically. We chose high values for these two hyperparameters to avoid overfitting on synthetic data. We use a dropout rate of 0.6, also found empirically, and a batch size of 10, which is the maximum we can fit on one GPU. Models are trained using Digital Research Alliance of Canada's Narval computing cluster on four Nvidia A100 GPUs, using 20 CPU cores and 100 GB of RAM. The training procedure completes in approximately 32 hours for all models. We train for 80,000 steps and select the model with the lowest Jensen-Shannon divergence between the validation set label distribution and the distribution of all our predicted labels. We also report metrics such as $R²$ and \gls{rmse}, but we decide to use the Jensen-Shannon divergence as the selection criterion, as we notice a tendency for predictions to concentrate at the extreme ends of the motion range. The Jensen-Shannon divergence helps us select the model that best represents the overall motion distribution.

\section{Results}

Our experiments aim to (1)assess our model's performance for motion prediction on both synthetic and real data; and (2) validate this estimated motion parameter  by evaluating previously reported associations between motion and cortical thickness and age.

\subsection{Motion Prediction}
\subsubsection{Performance on Synthetic Data}
\begin{figure}
    \centering
    \includegraphics[width=0.7\linewidth]{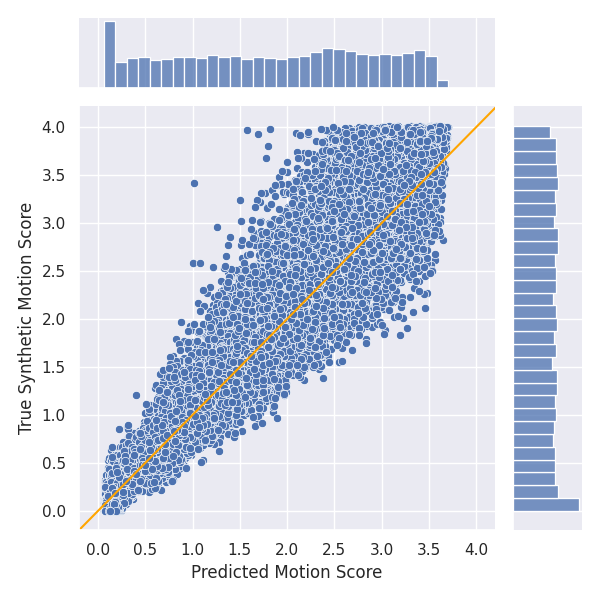}
    \caption{Prediction on the synthetic test set for our model compared to ground truth. The orange line represents $y = x$: the perfect prediction.}
    \label{fig:results:synth:pred}
\end{figure}

We first assess the performance of our method in the synthetic test set. Our model achieves an \gls{rmse} of 0.287 and a high $R^2$ of 0.94.  
In the prediction plot (Figure \ref{fig:results:synth:pred}), we notice that our predicted motion values are concentrated near zero, with no values exceeding approximately 3.5. We also observe greater deviation from the $y = x$ line as the motion increases. This effect can be explained by the nature of our task: we quantify actual simulated motion rather than the effect of motion on the volume. A single motion score can correspond to different expressions of artifacts, which become harder to distinguish as the motion level increases.

\subsubsection{Comparison with Real Motion}

We then test our model's ability to directly quantify motion on real artifacts using the three-level motion scores provided with the \gls{mrart} dataset. We compute the Spearman rank correlation between our continuous predictions and \gls{mrart}'s visual assessment grades. Our model achieves a high correlation of 0.71. This result demonstrates our model's capability for real motion prediction without fine-tuning on independent data.

\subsubsection{Ablation Study}
\begin{table}[]
\centering
    \begin{tabular}{lcc|cc|c}
    \toprule
         \multirow{2}{*}{Conv.} & \multicolumn{2}{c|}{Augmentation} & \multicolumn{2}{c|}{Synthetic Test Set}  & Spearman Rank\\
         & Base & Contrast & R² Score & RMSE & MR-ART\\
         \midrule
        1 & \xmark & \xmark & 0.917 & 0.339 &  -0.10\\ 
        2 & \xmark & \xmark & \bfseries 0.940 & \bfseries 0.287  & 0.11\\ 
        2 & \cmark & \xmark & 0.921 & 0.330& 0.66\\ 
        2 & \cmark & \cmark & 0.898 & 0.375  & \bfseries 0.71\\ 
         \bottomrule
    \end{tabular}
    \caption{Results of variations of SFCN on synthetic test set and \gls{mrart} dataset. SFCN2 without any augmentation performs better on synthetic data, but SFCN2 using all augmentation strategies has a stronger correlation with real motion scores.}
    \label{tab:result:ablation}
    
\end{table}
To assess the benefits of our approach, we perform an ablation study starting from the base SFCN to our two-convolution SFCN with both augmentation types (Table \ref{tab:result:ablation}).  
The results indicate that the SFCN2 without any augmentation outperforms the other variations on the synthetic test set. We explain these results by a discrepancy between the train/validation and test sets when augmentation is applied. Indeed, our augmented networks are exposed to a different domain that does not match the test set domain, which could explain the underperformance compared to a network specifically trained on this domain. We understand that the added expressiveness given to SFCN2 is beneficial for this task.

\begin{figure}
    \centering
    \includegraphics[width=\linewidth]{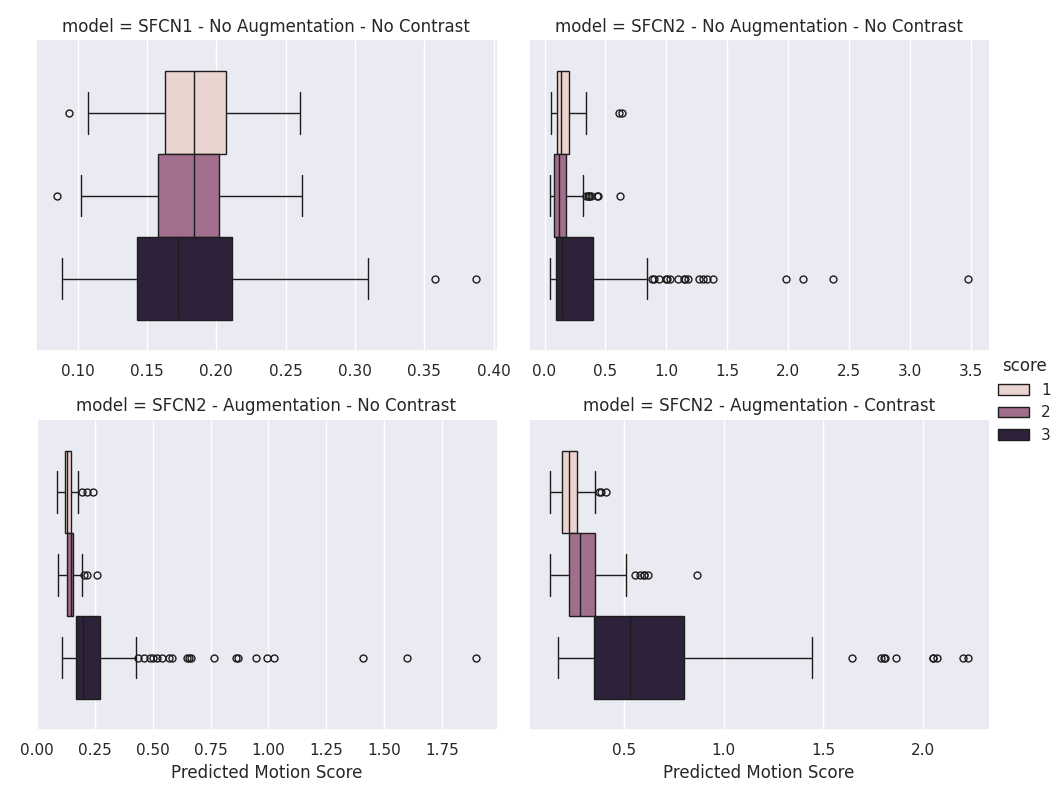}
    \caption{Predicted motion score distribution for each MR-ART motion label. The original SFCN (top left) shows no clear separation between motion classes. In contrast, the SFCN2 model with augmentations (bottom left and bottom right) demonstrates a clear ranking of motion scores, particularly when using contrast augmentation, indicating improved class separability.}
    \label{fig:results:mrart:boxplot}
\end{figure}

On the \gls{mrart} manual labels, the complete model outperforms the other variations by a large margin. This confirms that our model is more robust in real scenarios, even if its performance on synthetic data is lower. It is also interesting to note that the introduction of Gaussian noise and smoothing (base augmentation) greatly improves our motion regression capability to accurately rank the \gls{mrart} labels. These two augmentation steps appear to make our model more robust to real and unseen data. Visually, we also observe that our model's predictions follow the \gls{mrart} grading (Figure \ref{fig:results:mrart:boxplot}). A label of 1 corresponds to volumes with nearly no artifacts, whereas a label of 3 encompasses all magnitudes of severe artifacts. We can see this difference in range in the interquartile distances of our boxplots. Finally, we also visualize the uncertainty of intermediate classes, as the label 2 grade is not as distinctly separated.

\subsection{Motion Related Statistical Biases}
\subsubsection{Biases in Synthetic Motion}
We first want to assess whether synthetic motion impacts cortical thickness estimation as previously reported in studies with real data \citep{reuter_head_2015, alexander-bloch_subtle_2016, madan_age_2018}.

\begin{figure}[h]
    \centering
    \includegraphics[width=\linewidth]{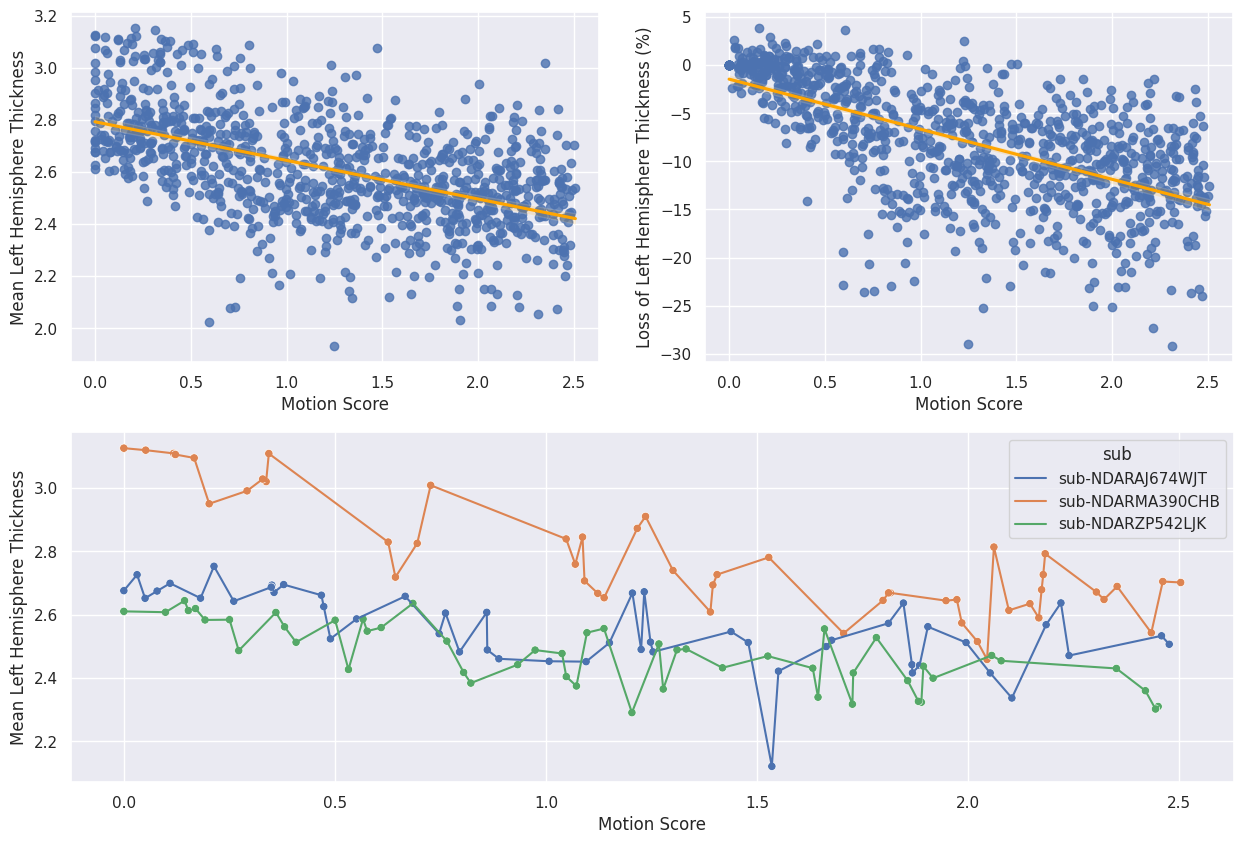}
    \caption{First row  Linear regressions of order one between Motion Score and Left Hemisphere Thickness (left) and percent loss in Left Hemisphere Thickness (right). Second row: Evolution of cortical thickness as we generate an increasing amount of motion on the same volumes}
    \label{fig:synthmotion:motionvsthick}
\end{figure}

We start by plotting the cortical thickness of the left hemisphere as a function of the motion score (Figure \ref{fig:synthmotion:motionvsthick}). We also compute the percent difference between the original thickness and the thickness of the corrupted scan by subtracting the generated scan thickness from the original scan thickness.
$$Loss(\text{original}, \text{synthetic}) = \frac{Thickness_{\text{synthetic}} - Thickness_{\text{original}}}{Thickness_{\text{original}}} \times 100$$

We then fit a \gls{glm} for the following relationship:
\begin{equation}\label{eq:stat:model}
    \text{Mean Left Hemisphere Cortical Thickness} = \beta_1 \cdot \text{age} + \beta_2 \cdot \text{sex} + \beta_3 \cdot \text{motion} + c
\end{equation}

We obtain impressive results (Table \ref{tab:synthmotion:GLMResults}) with pseudo-$R^2$, as well as a very low p-value, indicating a highly significant correlation between our motion score and FreeSurfer's cortical thickness. As this negative correlation is known to occur with real data, finding the same kind of relationship validates the assumption that training on synthetic artifacts may help estimate this bias with real volumes.

\begin{table}[h]
    \centering
    \begin{tabular}{lccccc}
    \toprule
                    & Coefficient ($\beta_i$) & Standard Error & P $> |$z$|$ & [0.025 & 0.975] \\
    \midrule
    Intercept    & 3.1017  & 0.027  & 0.000 & 3.049  & 3.155 \\
    Age          & -0.0229 & 0.002  & 0.000 & -0.026 & -0.020 \\
    Sex          & 0.0345  & 0.011  & 0.001 & 0.013  & 0.056 \\
    Motion Score & -0.1480 & 0.006  & 0.000 & -0.161 & -0.135 \\
    \bottomrule
    \end{tabular}
    \caption{\gls{glm} regression results. We obtain a Pseudo-$R^2$ of 0.583 and a p-value of $2e^{-116}$ for the motion parameter.}
    \label{tab:synthmotion:GLMResults}
\end{table}

\subsubsection{Correlation Between Estimated Motion and Cortical Thickness in Real Data}

\begin{figure}
    \centering
    \includegraphics[width=\linewidth]{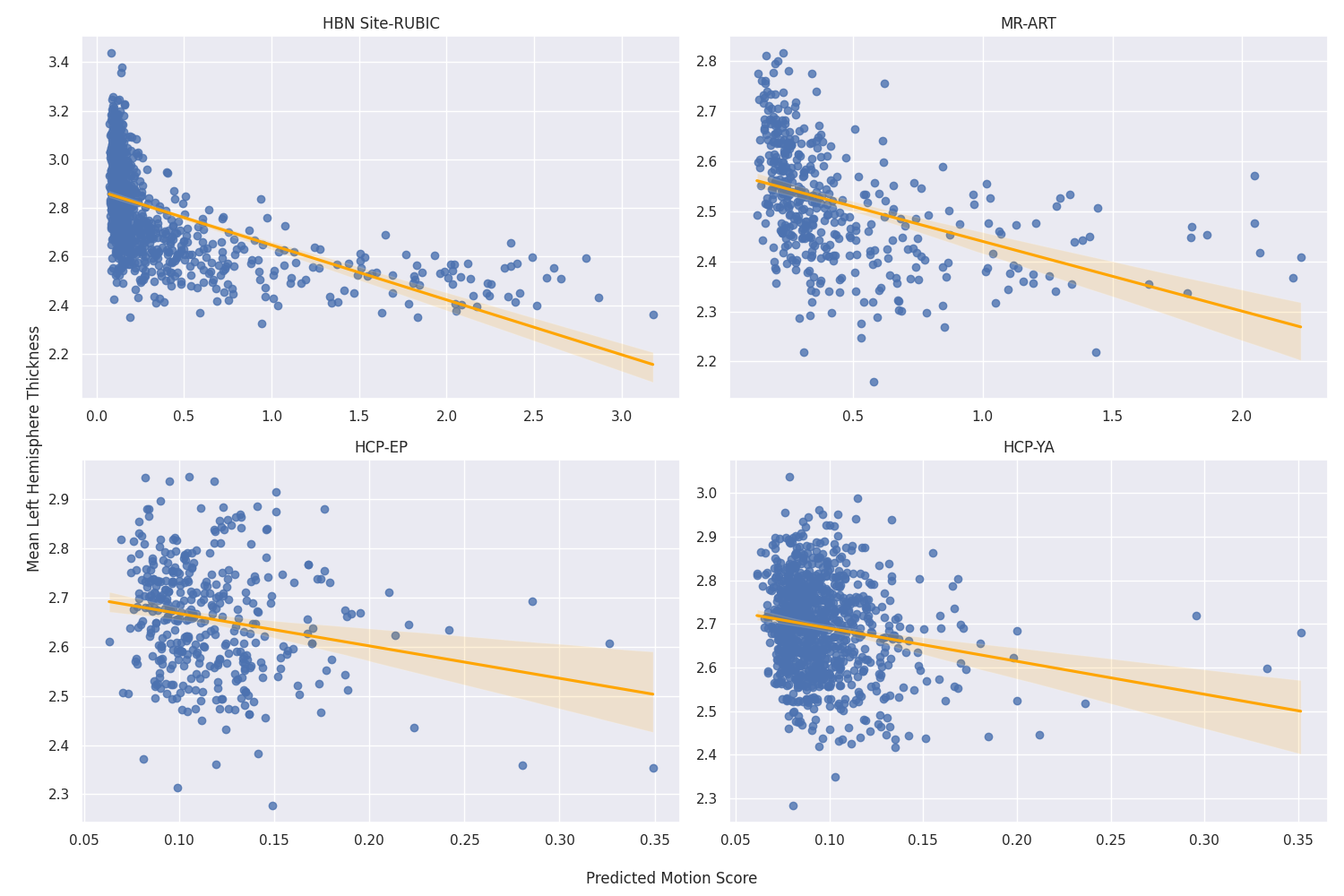}
    \caption{Regression between mean left hemisphere cortical thickness and our model's predicted motion score. There is a clear negative correlation for \gls{hbn} and \gls{mrart} (top left and top right). In contrast, the correlation for \gls{hcpep} and \gls{hcpya} (bottom left and top bottom) is weaker} 
    \label{fig:results:4dataset:reg}
\end{figure}

\begin{figure}
    \centering
    \includegraphics[width=\linewidth]{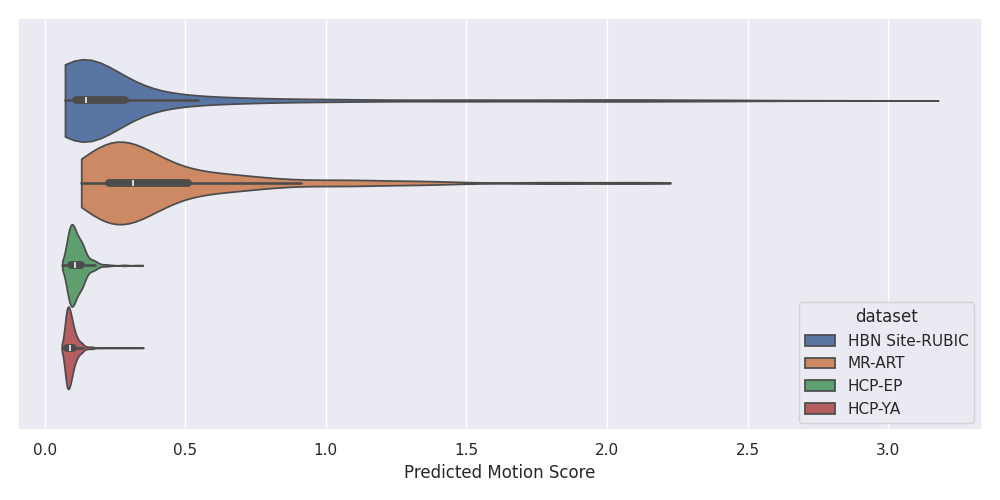}
    \caption{Predicted motion score distribution for each dataset.}
    \label{fig:results:4dataset:distrib}
\end{figure}

We start by qualitatively visualizing the relationship between motion and cortical thickness (Figure \ref{fig:results:4dataset:reg}).  
While we can see a clear downward trend in \gls{hbn} and \gls{mrart}, we notice that plots from \gls{hcp} show wider uncertainty toward larger motion scores. It is worth mentioning that these plots are a simplified visualization of the final relationship, which also accounts for age and sex. We also notice a stark difference in motion distribution (Figure \ref{fig:results:4dataset:distrib}); this makes sense given that \gls{hbn} subjects are children to young adults—a population more prone to movement—while MR-ART's \gls{mri} scans are artificially motion-corrupted. In contrast, \gls{hcp}'s strict quality standards and more mature cohorts should exhibit weaker motion artifacts.

We then assess these correlations from a quantitative standpoint by fitting the statistical model specified by Eq. \ref{eq:stat:model} for each dataset identified from OpenNeuro and the previously mentioned datasets. We report p-values and coefficients for the motion parameter (Table \ref{tab:openneuro:model}). Moreover, previous studies have shown that some structures are more impacted by motion than others. In addition to the mean thickness, we also fit a model for each of the 34 structures' thicknesses reported by FreeSurfer. As we test 15 different datasets, we obtain $15\times35 = 525$ models; hence, we apply Benjamini-Hochberg's \gls{fdr} correction \citep{benjamini_controlling_1995} to the full set of p-values to correct for accidental significance.

We obtain significant relationships in 12 of the15 datasets, which further indicates strong generalizability of our model. Furthermore, analyzing the motion coefficients for significant datasets, we find an average value of -0.208 $\pm$ 0.088, which agrees with the negative relationship observed in the literature and our synthetic data study. Moreover, we obtain a strong correlation with ds000144 and ds005234, showing robustness to different hardware.

With respect to our analysis of individual structures, we find that at least one structure is impacted by motion in every dataset. Given a conservative correction of p-values, this shows that our model detects a bias in every dataset.

\begin{table}[]
    \centering
    \begin{tabular}{lcccc}
\toprule
\multirow{2}{*}{Dataset} & \multicolumn{3}{c}{Motion's Impact on Mean Thickness} & Percentage of \\
                       & Coefficient & P-Value & \gls{fdr} corrected P-Value & Significant Structures \\
\midrule
HBN Site-RUBIC & -0.246 & 1.55e-143 &\bfseries 2.04e-141 & 97.14\% \\
HCP-YA & -0.651 & 2.73e-07 &\bfseries 1.75e-06 & 77.14\% \\
HCPEP & -0.904 & 1.21e-03 & \bfseries3.64e-03 & 51.43\% \\
MR-ART & -0.124 & 1.19e-21 &\bfseries 1.56e-20 & 85.71\% \\
ds000115 & -0.232 & 1.98e-04 & \bfseries7.42e-04 & 54.29\% \\
ds000144 & -0.301 & 3.72e-07 & \bfseries2.32e-06 & 51.43\% \\
ds000256 & -0.173 & 8.56e-05 & \bfseries3.60e-04 & 54.29\% \\
ds001486 & -0.100 & 3.42e-07 & \bfseries2.16e-06 & 77.14\% \\
ds001748 & -0.068 & 2.57e-01 & 3.47e-01 & 17.14\% \\
ds002424 & -0.213 & 5.46e-07 & \bfseries3.22e-06 & 74.29\% \\
ds002862 & -0.229 & 3.64e-03 &\bfseries 9.00e-03 & 40.00\% \\
ds002886 & -0.082 & 1.66e-02 & \bfseries3.43e-02 & 17.14\% \\
ds003499 & 0.075 & 6.80e-01 & 7.46e-01 & 2.86\% \\
ds003568 & -0.366 & 1.73e-01 & 2.49e-01 & 11.43\% \\
ds005234 & -0.332 & 2.63e-11 &\bfseries 2.51e-10 & 68.57\% \\
\bottomrule
\end{tabular}
    \caption{Results of fitting eq. \ref{eq:stat:model} for each dataset. P-values are adjusted with false discovery rate correction; bold p-values indicate $<$ 0.05}
    \label{tab:openneuro:model}
\end{table}

\subsubsection{Correlation Between Estimated Motion and Age}

As stated previously, multiple studies observe higher motion in younger and older subjects compare to middle aged individuals. To further validate our models, we test for the same type of relationship by fitting a \gls{glm} for the relationship:
\begin{equation}
    \text{motion} \sim \beta_1 \cdot \text{age}
\end{equation}
We test this relationship in \gls{hbn}'s RUBIC site, \gls{mrart}, and on the aggregation of all OpenNeuro data. We exclude \gls{hcpya} and \gls{hcpep} as their age range was between 18 and 36 years old, which is not reported to be strongly correlated with motion \cite{pardoe_motion_2016}. We find significant relationships for all datasets (Table \ref{tab:age_study:all_data}). It is worth noting that we obtain a negative coefficient for OpenNeuro and \gls{hbn} Site-RUBIC, which have age ranges of five–35 and five–21, respectively, whereas we find a positive coefficient for \gls{mrart}, which has subjects between 18 and 75. These findings agree are in line with the literature.

\begin{table}[]
    \centering
   \begin{tabular}{lcccc}
\toprule
\multirow{2}{*}{Dataset} & \multicolumn{4}{c}{Age} \\
                        & Coefficient & P-Value & \gls{fdr} Corrected P-Value & Range \\
\midrule
HBN Site-RUBIC     & -0.022 & 9.78e-08 & \bfseries 2.93e-07 & 5--21 \\
MR-ART             & 0.004  & 1.86e-03 & \bfseries 1.86e-03 & 18--75 \\
All OpenNeuro data & -0.010 & 1.14e-06 & \bfseries 1.71e-06 & 5--35 \\
\bottomrule
\end{tabular}
    \caption{Relationship between motion predicted by our best model and subject age.}
    \label{tab:age_study:all_data}
\end{table}

\section{Discussion}

In this work, we leverage synthetic motion augmentation to generate a large labeled dataset using objective quantification based on simulated patient motion. We show that training our \gls{sfcn}2 model on this synthetic dataset yields a high test $R^2$, demonstrating the ability of our architecture to learn synthetic motion. Our model's correlation with MR-ART's labels also indicates good prediction of real motion without fine-tuning on real artifacts.

We also study the impacts of our improvements compared to the base \gls{sfcn}. First, increasing the number of convolutions per \gls{sfcn} block to two yields significant improvements on the synthetic test set and \gls{mrart}'s manual labels. We also test two augmentation strategies: one leveraging Gaussian noise and smoothing, and another applying various contrast modification techniques. Both lead to increased Spearman rank correlation with \gls{mrart}'s motion labels. We believe the first strategy helps by creating more realistic volumes that combine fine motion information with additional noise, while the second strategy exposes our model to a variety of contrasts, effectively simulating different acquisition settings. On synthetic data, we notice that models using augmentation achieve lower $R^2$ and \gls{rmse} on the test set, which can be explained by a domain shift effect. Ultimately, our best-performing model on real data is the one incorporating all our modifications, demonstrating strong generalizability.

We then investigate the ability of our model to estimate known biases.  
First, we study how synthetic motion impacts the automatic measurement of cortical thickness and found a highly significant negative correlation, which is expected for real motion. This confirms that synthetic motion simulates not only direct artifacts but also other motion-related effects such as cortical thickness biases.  
To evaluate the relationship between cortical thickness and motion in real data, we visually show a clear relationship in \gls{hbn} and \gls{mrart}, while results are more uncertain for \gls{hcpya} and \gls{hcpep}. We explain this by the high quality standards of the two \gls{hcp} studies, which are reflected in the overall motion distribution across the four datasets. This supports the adequacy of our model, as it displays distributions that match prior knowledge about data quality.

Furthermore, we test our best model on 15 different datasets in total—11 from OpenNeuro—and obtain significant correlations between mean left hemisphere cortical thickness and predicted motion scores for 12 of them. We observe higher p-values for \gls{hcpya} and \gls{hcpep}, which can be explained by domain shift and the known tendency for reduced motion variance in young adults, combined with the high quality standards enforced by \gls{hcp}. Since OpenNeuro datasets involve fewer subjects and varying hardware/software, these results demonstrate our model's robustness to new data. Notably, we observe a significant relationship in a dataset acquired on a scanner from a different manufacturer.
When analyzing finer brain structures, we always find at least one region significantly correlated with our model's predictions. While it is known that motion affects cortical thickness differently depending on the structure, and even though we corrected our findings for multiple comparisons—it is likely that some of these findings are incidental. For example, we find only one affected structure out of 35 in dataset ds003499.

Finally, we test the relationship between age and motion, which has also been demonstrated in the literature, and found similar results. We observe significantly higher amounts of motion in both younger and older subjects.

\section{Conclusion}

This is the first attempt at correcting motion-related biases in automated anatomical measurements using a deep learning estimation of real patient motion through synthetic data. Our model is robust to variation in \gls{mri} hardware and software by comparing it to 15 datasets, unrelated to our training data. We also obtain good correlation with manually labeled motion scores on \gls{mrart}. A relationship between age and motion is similarly studied, and our results are in agreement with previous research. We conclude that our model learns, from purely synthetic motion artifacts, a regression that can be readily applied to \gls{mri} studies using sequences close to MPRAGE, without prospective motion correction. This allows motion to be included as a variable in statistical analyses of population studies. Including motion is important, especially for groups that tend to move more during scans, as motion can bias anatomical measurements. Our model provides a simple and reliable summary scalar that can support such analyses.

Future research should focus on testing this method on different scanners and studying how it could be adapted for volumes using prospective motion correction. It would also be interesting to expand this method to other kinds of artifacts that can be simulated, to design more accurate artifact simulators, and to use a broader set of regression metrics.

\section*{Data and Code Availability}
All data used comes from public or available under verification datasets. 
\begin{itemize}
    \item \gls{hbn} : \url{https://fcon_1000.projects.nitrc.org/indi/cmi_healthy_brain_network/MRI_EEG.html}
    \item \gls{hcpya} : \url{https://www.humanconnectome.org/study/hcp-young-adult/overview} (Require clearance)
    \item \gls{hcpep} :\url{https://www.humanconnectome.org/study/human-connectome-project-for-early-psychosis} (Require clearance)
    \item \gls{mrart} : \url{https://doi.org/10.18112/openneuro.ds004173.v1.0.2}
    \item ds000115 : \url{https://openneuro.org/datasets/ds000115/versions/00001}
    \item ds000144 : \url{https://openneuro.org/datasets/ds000144/versions/00002} 
    \item ds000256 : \url{https://openneuro.org/datasets/ds000256/versions/00002}
    \item ds001486 : \url{https://doi.org/10.18112/openneuro.ds001486.v1.3.1}
    \item ds001748 : \url{https://doi.org/10.18112/openneuro.ds001748.v1.0.4}
    \item ds002424 : \url{https://doi.org/10.18112/openneuro.ds002424.v1.2.0}
    \item ds002862 : \url{https://github.com/OpenNeuroDatasets-JSONLD/ds002862} (This dataset is indexed by OpenNeuro's NeuroBagel instance but is not accessible on OpenNeuro's website)
    \item ds002886 : \url{https://doi.org/10.18112/openneuro.ds002886.v1.1.0}
    \item ds003499 : \url{https://doi.org/10.18112/openneuro.ds003499.v1.0.1}
    \item ds003568 : \url{https://doi.org/10.18112/openneuro.ds003568.v1.0.4}
    \item ds005234 : \url{https://doi.org/10.18112/openneuro.ds005234.v2.1.7}
    
\end{itemize}

To ensure reproducibility, our manual labels, datasets splits, summary of freesurfer metrics, test predictions, and research code are available on GitHub at :  \url{https://github.com/Neuro-iX/cortical-motion}. We only excluded \gls{hcpep} and \gls{hcpya} as they require special access. We also provide our data review tool, \textit{MotionScore}, at : \url{https://github.com/Neuro-iX/MotionScore}.  

We designed a CLI tool and a python library called \textit{Agitation} :  \url{https://github.com/Neuro-iX/agitation}. This tool is available in PyPI's repositories, we also provide configuration files and a Docker container to use it as a \textit{Boutiques} command and a \textit{Nipoppy} pipeline : \url{https://doi.org/10.5281/zenodo.15346973}.
We also provide our models' weights on Zenodo : \url{https://doi.org/10.5281/zenodo.15288225}.

\section*{Author Contributions}
Charles Bricout wrote the code for all experiments. He also drafted the manuscript and contributed to developing the idea.
Samira Ebrahimi Kahou contributed to the design of machine learning architectures, editing and supervision.
Sylvain Bouix contributed to defining the research question, editing and supervision.



\section*{Declaration of Competing Interests}

We do not have any competing interests.



\printbibliography
\newpage
\appendix
\section{Motion Score Examples}

\begin{figure}
    \centering
    \includegraphics[width=0.75\linewidth]{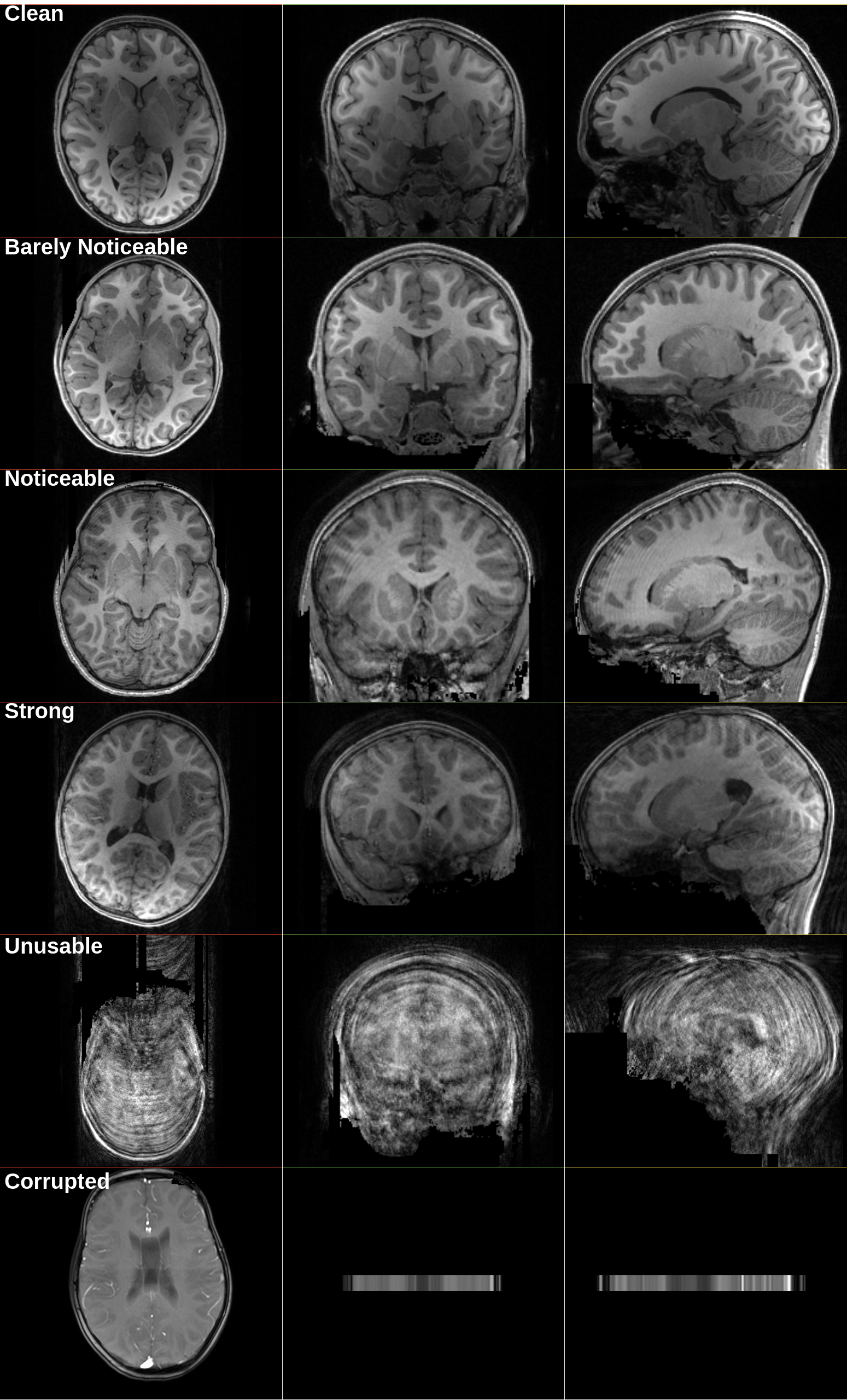}
    \caption{Examples from \gls{hbn} presented to Rater Two and Three}
    \label{fig:appendix:motsample}
\end{figure}

\end{document}